\documentclass[useAMS,usegraphicx]{mn2e}
\usepackage{psfig}
\title[Optical monitoring of the $z$=4.40 quasar Q\,2203$+$292]{Optical
monitoring of the $z$=4.40 quasar Q\,2203$+$292}
\author[Ovcharov et al.]
{E. P. Ovcharov$^{1}$\thanks{E-mail: evgeni@phys.uni-sofia.bg},
P. L. Nedialkov$^{1}$, A. T. Valcheva$^{4}$, V. D. Ivanov$^{3}$,
\newauthor
N. A. Tikhonov$^{2}$, I. S. Stanev$^{1}$, A. B. Kostov$^{4}$ and Ts. B. Georgiev$^{4}$
\\
\\
$^{1}$Department of Astronomy, University of Sofia, 5 James
Bourchier, Sofia 1164, Bulgaria\\
$^{2}$Special Astrophysical Observatory, Russian Academy of Sciences, N. Arkhyz, KChR 369167, Russia\\
$^{3}$European Southern Observatory, Ave. Alonso de Cordova 3107,
Casilla 19, Santiago 19001, Chile\\
$^{4}$Institute of Astronomy, Sofia 1784, Bulgaria}

\begin{document}


\pagerange{\pageref{firstpage}--\pageref{lastpage}} \pubyear{2002}

\maketitle

\label{firstpage}

\begin{abstract}
We report Cousins $R$-band monitoring of the high-redshift
($z$=4.40) radio quiet quasar Q\,2203+292 from May 1999 to
October 2007. The quasar shows maximum
peak-to-peak light curve amplitude of $\sim$0.3 mag during
the time of our monitoring, and $\sim$0.9 mag when combined
with older literature data. The rms of a fit to the light
curve with a constant is 0.08 mag and 0.2 mag, respectively.
The detected changes are at $\sim$3-sigma level. The quasar
was in a stable state during the recent years and it might
have undergone a brightening event in the past.
The structure function analysis concluded that the object
shows variability properties similar to those of the lower
redshift quasars.
We set a lower limit to the Q\,2203+292 broad line region mass
of 0.3-0.4 M$_\odot$.
Narrow-band imaging search for redshifted Ly$\alpha$ from other
emission line objects at the same redshift shows no emission
line objects in the quasar vicinity.
\end{abstract}


\section{Introduction}

Many quasars show short-term or/and long-term variability
(Ulrich, Maraschi \& Urry 1997). These changes in the source flux
help to constrain the physics and the size of the central engine.
Astronomers came to an early understanding that the central engines
of the AGNs and QSOs can not be resolved easily, if at all, but
that the variability timescales measure the size of the emitting
regions (i.e. Blandford \& McKee 1982). Later on, this became the
basis of reverberation studies (for a recent review see Peterson
et al. 2004 and the references therein) that could map the
innermost broad line regions (hereafter BLR). Nevertheless, the
exact variability mechanisms remain unclear.

Typically, the variability is aperiodic, but it does show some
dependencies on time lag, luminosity, wavelengths and redshift.
For example, the variability amplitude increases with time lag
(Vanden Berk et al. 2004), more luminous quasars are less variable
(Vanden Berk et al. 2004; de Vries et al. 2005; Giveon et al.
1999), and the variability increases toward the blue part of the
spectrum (Vanden Berk et al. 2004; de Vries et al. 2005). It is
particularly difficult to constrain the variability versus redshift
dependence because of the inevitable biases at high redshift,
limiting the quasar luminosity range, the probed time-baseline,
etc. of the more distant quasars.

The radio properties of QSOs also seem to be related to the
optical variability: the radio-loud ones are relatively more
variable that the radio quiet ones, and the blazars show even
stronger variability because of beaming, which is quite
different than the long-term variability studied in this work.
For a more comprehensive review of the QSO variability properties
we refer the reader to the summary of Wold, Brotherton \& Shang
(2007). As of now, there is no commonly accepted theory of the
QSO variability and the existence of some strange objects,
albeit rare ones, complicates the picture even further. A good
example is the radio-quiet QSO SDSS\,J153259.96-003944.1
(Stalin \& Srianand 2005) who is the prototype of the rare class
of weak (or absent) emission-line quasars (WLQs). This object
shows strong long-term variability and flat optical spectrum
like BL\,Lacs but there is no radio emission, no optical
polarization and no X-ray (Shemmer et al. 2006). These
properties may be explained by a deficit of line-emitting gas
in the vicinity of the central continuum source, similar to
the X-ray weak quasar PHL\,1811 (Leighly et al. 2007 and the
references therein).

The most comprehensive quasar variability studies, especially the
ones covering long time scales, are focused either on bright
low-redshift ones (Hook et al. 1994; Hawkins 2002), or they
study the behavior of the structure function for large samples
(de Vries, Becker \& White 2003; de Vries et al.
2005; Hovatta et al. 2007). The literature is lacking well sampled
light curves of individual distant quasars, with the notable
exception of Kaspi et al. (2007) with whom we share five objects
from our extended program.
While challenging, the distant QSOs are also rewarding because the
variability can help constrain the size of the QSO accretion
disks (Hawkins 2007) at early times. Furthermore, identifying
variable high-redshift quasars is a necessary first step for
reverberation mapping of their broad line regions (i.e. Kaspi et
al. 2007). These arguments motivated us to start an optical
monitoring study of QSOs with $z$$\geq$4.

Here we present a photometric sequence for Q\,2203+292, which is one
of the first detected quasars within that redshift range (Dickinson \&
McCarthy 1987) and it is radio-quiet (Schneider et al. 1992; Schmidt
et al. 1995; Omont et al. 1996). Surdej et al. (1993), Crampton,
McClure \& Fletcher (1992) and Kochanek (1993) reported that it is
not gravitationally lensed.
The only study in the literature on the Q\,2203+292 variability
comes from McCarthy et al. (1988) who concluded from three Lick 3m
plates and two plates from the 5m Hale telescope (Longair \& Gunn
1975; Riley, Longair \& Gunn 1980) that the quasar did not vary
strongly over a period of the 15 years since its discovery. A
few additional broadband observations of Q\,2203+292 are reported
elsewhere, but they are in different photometric systems
(see Table~\ref{Table_Literature}).

Turner (1991) calculated for Q\,2203+292 a mass of the central
black hole of (0.69-9.6)$\times$10$^{8}$ M$_\odot$ and a mass
accretion rate of 1.6-22 M$_\odot$/yr, depending on the adopted
cosmological model. This values are typical for the quasars at
the same redshift (Turner 1991; Dietrich \& Hamann 2004).

\begin{table}
\caption{Broad band photometry of Q\,2203+292 since 1987.}
\label{Table_Literature}
\begin{tabular}{c@{  }c@{ }c@{  }c} \hline
Date       & Reference                   & Filter        & Brightness     \\
yyyy/mm/dd &                             &               & mag            \\
\hline
1987/09/25 & McCarthy et al. (1988) & $r_{\rmn{s}}$ & 20.78$\pm$0.09 \\
           &                             & $V$           &  22.0$\pm$0.3  \\
           &                             & $I$           &  21.1$\pm$0.3  \\
           &                             &               &                \\
1988/09/12 & Schneider, Schmidt \& Gunn   & $r_{\rmn{4}}$ & 20.88$\pm$0.05 \\
           &         (1989)              & $g_{\rmn{4}}$ & 22.33$\pm$0.06 \\
           &                             &               &                \\
1989/09/09 & Crampton et al. (1992)      & $R_C$           &  20.7$\pm$0.1  \\
           &                             & $V$           &  22.0$\pm$0.1  \\
           &                             &               &                \\
1990/06/21 &    Crampton                 & $R_C$           &  20.4$\pm$0.1  \\
           &(private communication)      &               &                \\

\hline
\end{tabular}
\end{table}

\section[]{Observations and Data Reduction}

\subsection[]{Observing Strategy and Basic Data Reduction}

We monitored Q\,2203+292 in $R$-band with a variety of instruments
and telescopes. ESO Archive images were also used. The observing
log is shown in Table~\ref{Table_ObsLog}. Typically, the total
integration time was split into a few separate frames (as listed
in the last column) and the telescope was jittered by a few arcsec
between each of them to remove the artifacts caused by the
detector's cosmetic defects. All observations were performed in
clear, photometric nights. The object was monitored during
culmination, whenever possible, to minimize the airmass variation
during the observations.

\begin{table*}
\begin{center}
\begin{minipage}{150mm}
\caption[]{Observing log.}
\label{Table_ObsLog}
\begin{tabular}{c@{   }c@{   }c@{   }c@{   }c@{   }c@{   }c@{   }c@{   }c}
\hline
Data & Instrument@Telescope@Site & Pixel Scale & FoV & Airmass & FWHM & Total Integration Time \\
yyyy/mm/dd & & [arcsec\,px$^{-1}$] & [arcmin$^2$] & &[arcsec]&[s]\\
\hline
1999/05/15 & FORS1@VLT@Paranal              & 0.25 & 6.8$\times$6.8 & 1.76 & 0.4 & 1$\times$100=100 \\
2003/08/26 & Photometrics\,AT200A@2m@Rozhen & 0.29 & 5.0$\times$5.0 & 1.04 & 1.1 & 3$\times$1200=3600 \\
2004/10/09 & Photometrics\,AT200A@2m@Rozhen & 0.29 & 5.0$\times$5.0 & 1.03 & 1.4 & 3$\times$1200=3600 \\
2005/10/01 & EEV/Marconi\,42-40@1m@SAO      & 0.27 & 4.6$\times$4.6 & 1.23 & 1.2 & 3$\times$200=600 \\
2005/11/04 & VersArray\,1300B@2m@Rozhen     & 0.26 & 5.7$\times$5.5 & 1.11 & 1.8 & 9$\times$300=2700 \\
2005/11/06 & VersArray\,1300B@2m@Rozhen     & 0.26 & 5.7$\times$5.5 & 1.07 & 2.0 & 6$\times$600=3600 \\
2006/06/29 & Photometrics\,AT200A@2m@Rozhen & 0.29 & 5.0$\times$5.0 & 1.54 & 1.6 & 10$\times$180+2$\times$300=2400 \\
2006/08/18 & Photometrics\,AT200A@2m@Rozhen & 0.29 & 5.0$\times$5.0 & 1.04 & 1.4 & 3$\times$1200=3600 \\
2006/08/19 & Photometrics\,AT200A@2m@Rozhen & 0.29 & 5.0$\times$5.0 & 1.02 & 1.4 & 2$\times$1200=2400 \\
2006/08/25 & FoReRo2@2m@Rozhen              & 0.82 & 7.0$\times$7.0 & 1.32 & 2.4 & 8$\times$300=2400 \\
2006/10/23 & EEV/Marconi\,42-40@1m@SAO      & 0.27 & 4.6$\times$4.6 & 1.04 & 1.1 & 3$\times$300=900 \\
2006/10/24 & EEV/Marconi\,42-40@1m@SAO      & 0.27 & 4.6$\times$4.6 & 1.09 & 1.0 & 3$\times$300=900 \\
2007/09/10 & VersArray\,1300B@2m@Rozhen     & 0.26 & 5.7$\times$5.5 & 1.14 & 1.5 & 2$\times$1200=2400 \\
2007/10/04 & VersArray\,1300B@2m@Rozhen     & 0.26 & 5.7$\times$5.5 & 1.04 & 1.0 & 3$\times$1200=3600 \\
\hline
\end{tabular}
\end{minipage}
\end{center}
\end{table*}

\begin{figure}
\hspace*{0.0cm}\psfig{file=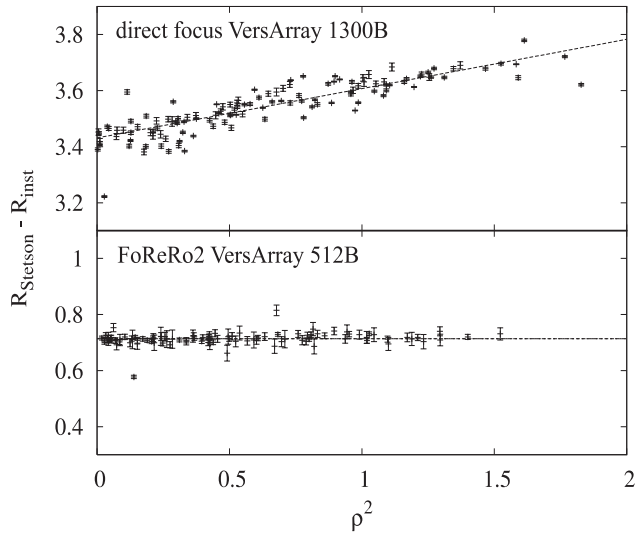,width=8.3cm,height=7cm}
\caption{Radial flux variation. The difference $R_{Stetson} -
R_{inst}$ magnitudes for all detectable stars in the field of the
globular clusters NGC\,7790 and NGC\,2420, taken with
VersArray\,1300B in direct focus (top) and VersArray\,512B in red
channel of FoReRo2 (bottom) versus the squared normalized distance
from the center of the detector $\rho^2$.}
\label{fig1}
\end{figure}

\begin{figure}
\hspace*{0cm}\psfig{file=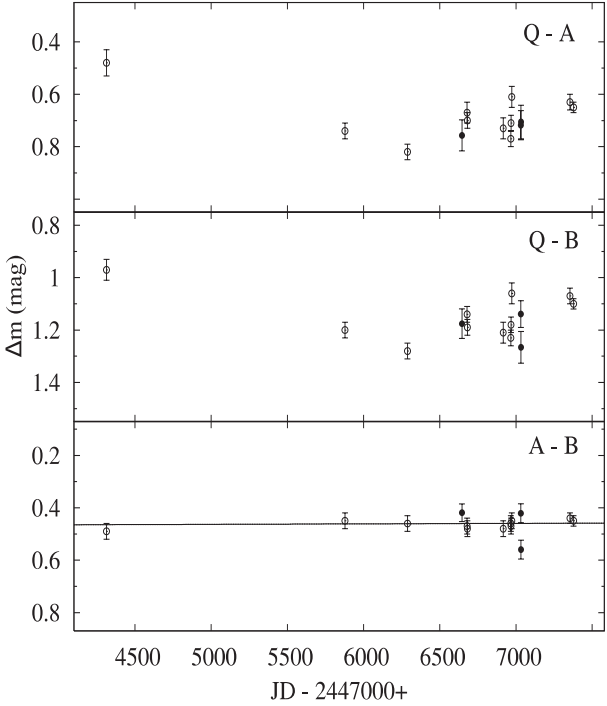,width=8.0cm,height=9.3cm}
\caption{Difference between the magnitudes of the quasar Q\,2203$+$292 and the
reference stars A and B is shown on the top and middle panels. The difference between
A and B reference stars is shown on the bottom panel. The solid line is the
linear fit to the data. The first open circle indicates the VLT data and
the other open circles are Rozhen Observatory data.
The solid circles are the data from the 1m SAO telescope.}
\label{fig2}
\end{figure}

\begin{figure*}
\hspace*{0.5cm}\psfig{file=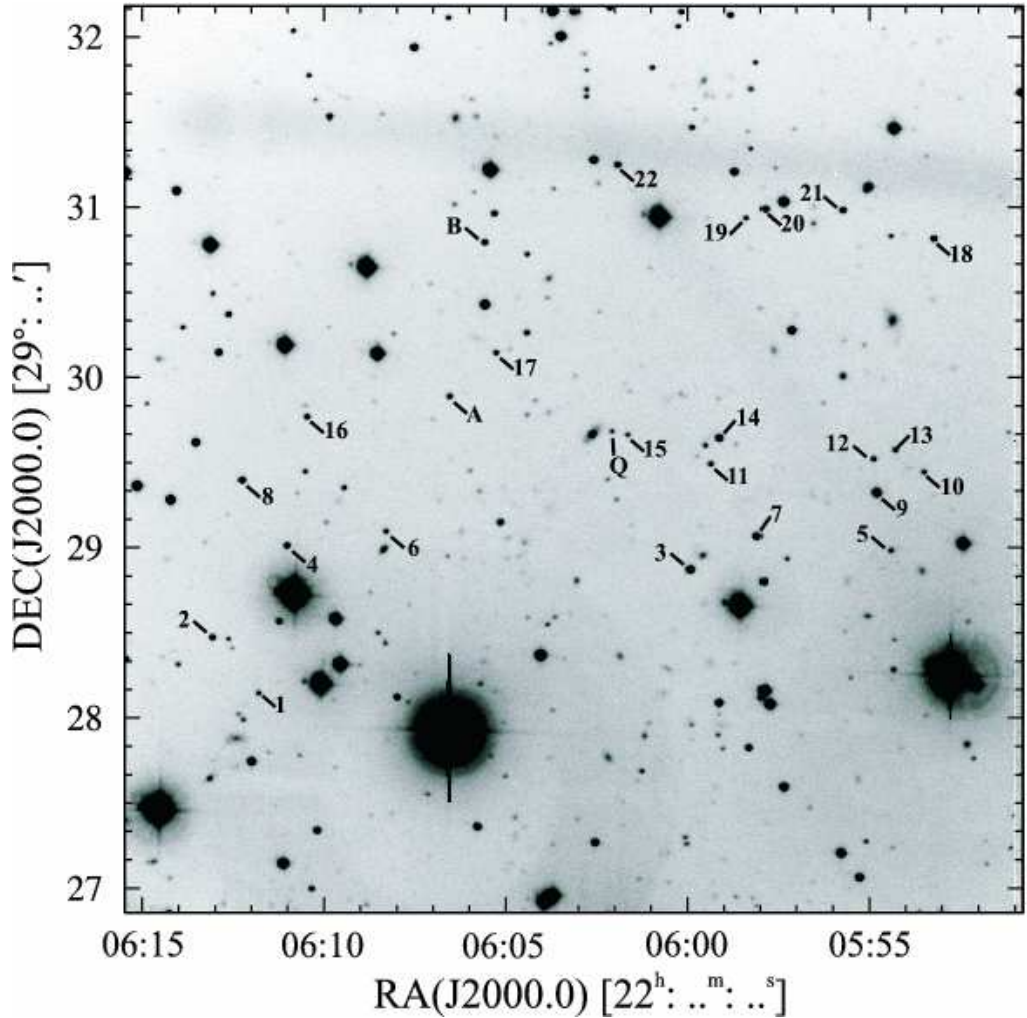,width=13.5cm,height=13.5cm}
\caption{Identification chart with standard stars in the field of
Q\,2203+292.
The image was obtained with 2m RCC telescope of NAO, Rozhen on
2004 October 9. The field of view is 5$\times$5 arcmin$^2$.
North is up, and East is to the left.}
\label{fig3}
\end{figure*}

The basic data reduction includes: bias substraction, flat fielding,
alignment of individual frames and combination. We used the standard
IRAF\footnote{IRAF
is the Image Reduction Analysis and Facility made available to the
astronomical community by the National Optical Astronomy Observatories,
which are operated by AURA, Inc., under contract with the U.S.
National Science Foundation. STSDAS is distributed by the Space
Telescope Science Institute, which is operated by the Association
of Universities for Research in Astronomy (AURA), Inc., under NASA
contract NAS5-26555.} routines to perform them.

\subsection[]{Instrumental Magnitudes and Corrections for Systematical Effects}

We carried out aperture photometry on the combined images. The
aperture diameter was set to the size of the FWHM to optimize the
signal-to-noise, because the sky contributed comparable flux to the
target flux in the wings of the image. This is possible because:
(i) the QSO PSF is indistinguishable from the PSF profiles of the
stars, and
(ii) the PSF variations across the field of view are negligible.
Therefore, the selection of the aperture does not have an effect on
the relative photometry we obtained on each individual image. Next,
the zero points we determined by comparing the instrumental and
the standard magnitudes of the calibration field (measured the same
way) contain into themselves the aperture corrections. For more
details on the photometric calibration see
Sec.~\ref{Sec_Standards}.

The Photometrics\,AT200A and the VersArray\,1300B cameras exhibit
spatial flux variations (Fig.~\ref{fig1}). They were removed as
described in Markov (2005), correcting the instrumental magnitudes
as follows:
\begin{equation}
R = c_{\rho} \rho^2 + R_{inst},
\end{equation}
where $c_{\rho}$ is a known coefficient (0.12 for
Photometrics\,AT200A and 0.17 for VersArray\,1300B) and $\rho$ is
the distance from the center of the detector. This is normalized
by the detector half-size, and it varies from $0$ to $\sqrt{2}$:
\begin{equation}
\rho = \sqrt{ \left(1-\frac{x}{x_c}\right)^2 +  \left(1-\frac{y}{y_c}\right)^2}.
\end{equation}
Here $x$ and $y$ are the coordinates of the object in pixels and
$x_c$ and $y_c$ are the coordinates of the central pixel. Note
that for the purpose of our differential photometry this
correction is minor, because the quasar and the comparison stars
were placed at the same position, within the pointing errors, so
$\Delta \rho^2$$<$0.1, translates into 0.02 mag extra uncertainty
in the differential magnitudes Q--A and Q--B (see Fig. 2). An
additional source of error, albeit also small, is the jittering
between the individual frames, that comprise the individual epochs
of our light curve. It was always 3-4 arcsec (except 12 arcsec in
one case) which corresponds to ignorable
magnitude error. Interestingly, the FoReRo2 (Jockers
et al. 2000) shows no spatial effects, as verified from observations
of Stetson standards (Stetson 2000) taken during a few different
photometric nights.

Differential light curves of the quasar were generated relative to two
nearby comparison stars imaged on the same frame. They were selected among
the most stable stars in the field (see Sec.~\ref{Sec_Standards}). The results
of the differential photometry are given in Table~\ref{Table_LightCurves} and
Fig.~\ref{fig2}. The points from SAO (solid circles) are with bigger errors,
because of the smaller telescope aperture and the shorter exposure times
(Table~\ref{Table_ObsLog}).

\begin{table}
\begin{center}
\caption{Differential instrumental magnitudes between the quasar and the
comparison stars A ($\alpha_{2000}$=22:06:07.15, $\delta_{2000}$=29:30:12.9)
and B ($\alpha_{2000}$=22:06:06.33, $\delta_{2000}$=29:31:08.0) and between
the two comparison stars.}
\label{Table_LightCurves}
\begin{tabular}{cccc}
\hline
Date       &     Q$-$A	   & Q$-$B	   &	 A$-$B      \\
yyyy/mm/dd &     [mag]	   & [mag]	   &	 [mag]      \\
\hline
1999/05/15 & 0.48$\pm$0.05 & 0.97$\pm$0.04 & 0.49$\pm$0.03 \\
2003/08/26 & 0.74$\pm$0.03 & 1.20$\pm$0.03 & 0.45$\pm$0.03 \\
2004/10/09 & 0.82$\pm$0.03 & 1.28$\pm$0.03 & 0.46$\pm$0.03 \\
2005/10/01 & 0.76$\pm$0.06 & 1.18$\pm$0.06 & 0.42$\pm$0.03 \\
2005/11/04 & 0.67$\pm$0.04 & 1.14$\pm$0.03 & 0.47$\pm$0.03 \\
2005/11/06 & 0.70$\pm$0.03 & 1.19$\pm$0.03 & 0.48$\pm$0.03 \\
2006/06/29 & 0.73$\pm$0.04 & 1.21$\pm$0.04 & 0.48$\pm$0.03 \\
2006/08/18 & 0.77$\pm$0.03 & 1.23$\pm$0.03 & 0.46$\pm$0.03 \\
2006/08/19 & 0.71$\pm$0.03 & 1.18$\pm$0.03 & 0.47$\pm$0.03 \\
2006/08/25 & 0.61$\pm$0.04 & 1.06$\pm$0.04 & 0.45$\pm$0.03 \\
2006/10/23 & 0.72$\pm$0.06 & 1.14$\pm$0.05 & 0.42$\pm$0.04 \\
2006/10/24 & 0.71$\pm$0.06 & 1.27$\pm$0.06 & 0.56$\pm$0.04 \\
2007/09/10 & 0.63$\pm$0.03 & 1.07$\pm$0.03 & 0.44$\pm$0.02 \\
2007/10/04 & 0.65$\pm$0.02 & 1.10$\pm$0.02 & 0.45$\pm$0.02 \\
\hline
\end{tabular}
\end{center}
\end{table}

\subsection[]{Photometric Calibration\label{Sec_Standards}}

The absolute calibration of our instrumental magnitudes includes
three steps. First, we tied the FORS1@VLT image to the Landolt (1992)
standard field MARK\,A. Then, we searched for all stars in common
between the FORS1@VLT image and eight other images from NAO Rozhen,
obtained under photometric conditions until 2006 August and calculated
transformations between the individual frames. Finally, using
these nine images we derived new magnitudes for a
few additional stars, bringing the number of reference stars
in the field to 24. Here we consider only stars with
rms$\leq$0.04 mag making sure the calibration is based only on
non-variable sources (Table~\ref{Table_Stdandards} and Fig.~\ref{fig3}).
The weights used for averaging the magnitudes were $\sigma^{-2}$,
where:
\begin{equation}
\sigma = \sqrt{\sigma_{zp}^{2} + \sigma_{inst}^{2}}.
\end{equation}
Here $\sigma_{zp}$ is the error of the zero-point and $\sigma_{inst}$
is the instrumental magnitude's error. The total uncertainty is
dominated by the zero-point errors with a typical value to
$\sim$0.03 mag. The instrumental error of individual measurements
attains 0.01 mag at $R_C$$\sim$20.5 mag level
(Table~\ref{Table_Stdandards}).

\begin{table}
\caption{Reference stars in the field of the quasar Q\,2203+292.
The quasar magnitude is for the 2004 October 9 image.}
\label{Table_Stdandards}
\begin{center}
\begin{tabular}{c@{  }c@{  }c@{   }c} \hline
 ID & \multicolumn{2}{c}{RA (2000.0) DEC} & R$_C$(err), mag \\
\hline
  Q & 22:06:02.70 & 29:30:02.0 & 20.51\,(0.03) \\
  A & 22:06:07.15 & 29:30:12.9 & 19.69\,(0.01) \\
  B & 22:06:06.33 & 29:31:08.0 & 19.23\,(0.02) \\
  1 & 22:06:12.17 & 29:28:26.0 & 20.29\,(0.03) \\
  2 & 22:06:13.47 & 29:28:45.3 & 19.43\,(0.03) \\
  3 & 22:06:00.42 & 29:29:13.7 & 18.58\,(0.03) \\
  4 & 22:06:11.51 & 29:29:17.9 & 19.40\,(0.04) \\
  5 & 22:05:54.95 & 29:29:22.6 & 20.23\,(0.03) \\
  6 & 22:06:08.80 & 29:29:24.1 & 20.03\,(0.03) \\
  7 & 22:05:58.62 & 29:29:26.2 & 18.60\,(0.02) \\
  8 & 22:06:12.79 & 29:29:40.8 & 19.03\,(0.02) \\
  9 & 22:05:55.39 & 29:29:42.9 & 17.72\,(0.03) \\
 10 & 22:05:54.13 & 29:29:50.9 & 20.26\,(0.03) \\
 11 & 22:05:59.95 & 29:29:51.6 & 19.96\,(0.03) \\
 12 & 22:05:55.51 & 29:29:55.0 & 19.99\,(0.04) \\
 13 & 22:05:54.92 & 29:29:58.5 & 19.88\,(0.03) \\
 14 & 22:05:59.73 & 29:30:00.8 & 18.61\,(0.03) \\
 15 & 22:06:02.24 & 29:30:01.1 & 20.45\,(0.03) \\
 16 & 22:06:11.07 & 29:30:04.1 & 19.67\,(0.03) \\
 17 & 22:06:05.93 & 29:30:29.0 & 19.91\,(0.03) \\
 18 & 22:05:54.04 & 29:31:13.9 & 19.19\,(0.03) \\
 19 & 22:05:59.18 & 29:31:19.4 & 20.18\,(0.03) \\
 20 & 22:05:58.68 & 29:31:22.7 & 18.80\,(0.02) \\
 21 & 22:05:56.55 & 29:31:23.1 & 19.08\,(0.02) \\
 22 & 22:06:02.73 & 29:31:37.0 & 19.24\,(0.02) \\
\hline
\end{tabular}
\end{center}
\end{table}

\subsection{Literature data and filter transformation\label{Sec_MagTransf}}

The observations of McCarthy et al. (1988), Schneider et al. (1989)
and (Crampton et al. 1992; private communication) provided additional
measurements of Q\,2203$+$292 (Table~\ref{Table_Literature}). Since
some of them were observed in filters other than the $R_C$ filter used
by us and we have to compare the luminosity of Q\,2203+292 with the
luminosities of quasars at similar redshift observed in $r^\prime$,
we were forced to derive transformations to the Cousins $R_C$
system.

We calculated $R_C$$-$$r$ for a $z$=4.4 QSO using the Q\,2203+292
spectrum from Constantin et al. (2002) convolving it with the
transmission curves of the standard $R_C$ filter and the other $r$
filters used in the literature studies. To obtain the necessary
spectral coverage, we combined it with the bluest part of the
Q\,2203+292 spectrum from McCarthy et al. (1988), shortwards of
1070 $\AA$.
Here and throughout this paper we used the following zero-point
fluxes: 3080 Jy for $R_C$ (Bessell 1979), 2810 Jy for $r_s$ system
(Djorgovski 1985), 4471 Jy for $r_4$ (Frei \& Gunn 1994) and
3631 Jy for $r^\prime$ AB system (Oke \& Gunn 1983).

\begin{figure}
\vspace*{0.cm}\psfig{file=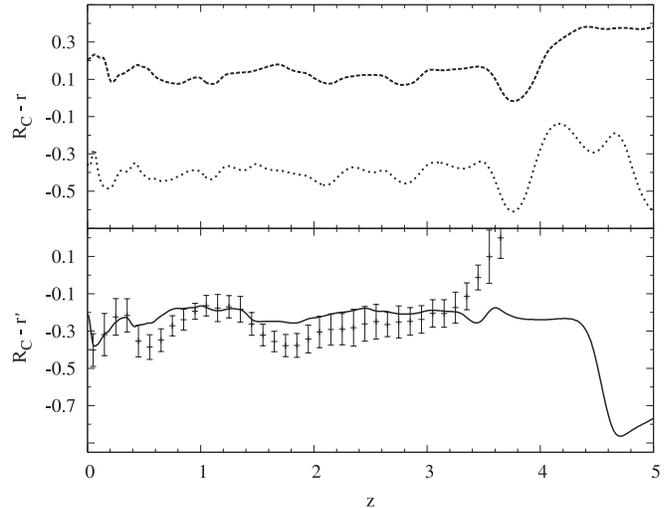,width=8.6cm,height=6.7cm}
\caption{The $R_C-r$ colours versus the redshift $z$. The upper panel
shows colour terms for $r_s$ (dashed line) and $r_4$ (dotted line). The
bottom panel shows $R_{C}-r^\prime$ (solid line) and the behavior of
the mean colour of the SDSS QSO sample according to Jester et al.
(2005). The synthetic colours were
derived from our composite QSO spectrum.}
\label{fig4}
\end{figure}

We also calculated $R_C$$-$$r$ as a function of redshift
$z$ to compare our corrections with the literature, and to derive
K-corrections for a comparison of the properties of Q\,2203+292 and
the SDSS QSOs (see Sec.~\ref{SecPhysProp}). We created our own
composite spectrum, combining the Constantin et al. (2002) median
spectrum with the bluest part of the McCarthy et al. (1988)
shortwards of 1150 $\AA$ (which contributes very little flux to
any of the $r$ filters). Finally, the reddest part of our template
($\lambda$$>$1450 $\AA$) comes from the Vanden Berk et al. (2001).
Indeed, the composite is dominated by lower redshift ($z$$\leq$2)
quasars. The overall behavior of $R_C$$-$$r$ as a function of $z$
is shown in Fig.~\ref{fig4}. The bottom panel demonstrates the
agreement between our synthetic photometry and the mean
$R_C$$-$$r^\prime$ colour for the quasars with $r^\prime$ photometry
from the SDSS QSO sample (Schneider et al. 2005) after applying
the colour equations of Jester et al. (2005; see their table 1).
Although these transformations were derived for quasars with
$z$$<$2.1, the figure suggests that they can be extrapolated even
up to $z$$\sim$3.3. For higher redshifts, Ly$\alpha$ enters the
filter passband rendering them unusable.

\begin{figure}
\hspace*{-0.2cm}\psfig{file=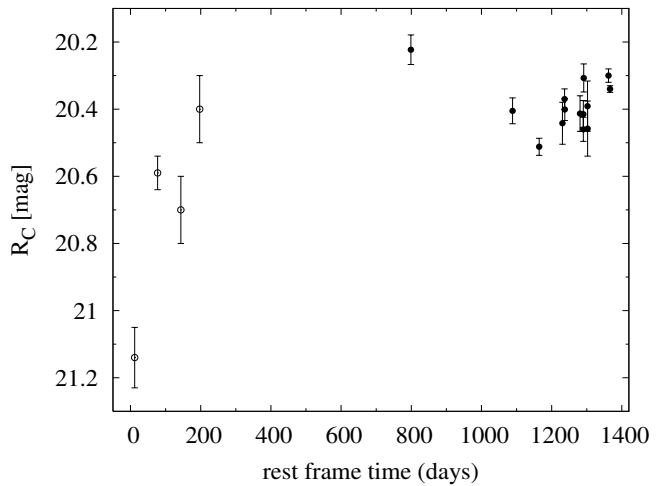,width=9.8cm,height=6.7cm}
\caption{Light curve of the quasar Q\,2203+292 in the rest frame. The open
circles show the literature data (Table 1) and the filled circles are our
measurements.}
\label{fig5}
\end{figure}

The corrections for Q\,2203+292 in $R_C$ system are:
$R_C$$-$$r_s$=+0.36 mag, $R_C$$-$$r_4$=$-$0.29 mag. We assign
to them tentative errors equal to the differences between the
$R$$-$$r$ values derived for this redshift from the quasar
spectrum and from our own composite spectrum: 0.02 mag for both
$R_C$$-$$r_s$ and $R_C$$-$$r_4$. They were added
in quadrature to the observational uncertainties of the first
two data points of our light curve (Fig.~\ref{fig5}).
Table~\ref{Table_UnifLightCurge} lists all available photometry
for Q\,2203+292 in the $R_C$ filter, including the corrected
literature measurements. Note, that here we combine systematic
with random errors, though.

\begin{table}
\begin{center}
\caption[]{Final $R_C$ light curve of Q\,2203+292.
See Sec.~\ref{Sec_MagTransf} for details. References:
1 McCarthy et al. (1988),
2 Schneider et al. (1989),
3 Crampton et al. (1992),
4 Crampton (private communication),
5 this work.}
\label{Table_UnifLightCurge}
\begin{tabular}{c@{   }c@{   }c@{   }c}
\hline
JD-2447000+& $R_C$(error),mag & Ref. \\
\hline
0064.04166 & 21.14 (0.09) & 1 \\
0417.28020 & 20.59 (0.05) & 2 \\
0773.84030 & 20.70 (0.10) & 3 \\
1064.04170 & 20.40 (0.10) & 4 \\
4313.43375 & 20.22 (0.04) & 5 \\
5878.38186 & 20.41 (0.04) & 5 \\
6288.31451 & 20.51 (0.03) & 5 \\
6645.39887 & 20.44 (0.06) & 5 \\
6679.30514 & 20.37 (0.03) & 5 \\
6681.28100 & 20.40 (0.03) & 5 \\
6916.41181 & 20.41 (0.05) & 5 \\
6966.39679 & 20.46 (0.04) & 5 \\
6967.42921 & 20.42 (0.04) & 5 \\
6972.54666 & 20.31 (0.04) & 5 \\
7032.18690 & 20.39 (0.08) & 5 \\
7033.28848 & 20.46 (0.08) & 5 \\
7354.46565 & 20.30 (0.02) & 5 \\
7378.27326 & 20.34 (0.01) & 5 \\
\hline
\end{tabular}
\end{center}
\end{table}

\section{Discussion}

\subsection[]{Variability}

The quasar shows (Fig.~\ref{fig5}) a brightness increase of $\sim$0.75 mag
at the beginning of the coverage but it is nearly constant later.
Due to the gaps in the lightcurve, any non-linear fluctuations (such
as flares) cannot be ruled out. We verify the variability properties 
of the Q\,2203+292 by means of differential photometry and Monte Carlo 
simulation.

The differential light curves (Fig.~\ref{fig2}) of the quasar were
generated with respect to the reference stars A and B, 
two of the most stable stars near the quasar.
The rms of the relative light curve A--B is 0.035 mag, and since
the two stars have similar magnitudes, their individual errors are
$\sim$0.025 mag. The rms for the quasar light curve with respect
to star A (marked as Q--A) is 0.083 and with respect to star B
(marked as Q--A) is 0.086 mag. The maximum peak-to-peak variation of
the quasar is $\sim$0.92 mag over the entire monitoring period
(Table.~\ref{Table_UnifLightCurge}) but it is reduced to 0.21 mag if
we consider only the Rozhen observations. The rms of all 18 QSO
measurements is 0.20 mag, and if only our 14 measurements are
considered, it decreases to 0.08 mag.

To test further the variability of Q\,2203+292, we carried out a
Monte Carlo simulation drawing 18 measurements from a constant source
with the measured mean magnitude of Q\,2203+292. Each of these points
was generated from a Gaussian distribution with the observational
error of the corresponding measurement, so that the artificial datasets
more faithfully represent the properties of the real observations. If
we consider all data, including the ones from the literature, none of
one million simulated data sets exceeded the observed rms However,
the colour transformations can be a source of extra uncertainty, so we
carried the same simulation only for our 14 measurements to obtained
that in 98.5 per cent of the cases the data are inconsistent with a
constant source. Excluding the VLT point lowers this probability down
to 86.4 per cent.

We conclude that if the colour transformation of the historical
observations can be considered reliable, the quasar have undergone a
brightening episode in the past but the unaccounted systematic effects
stop is from making a strong statement about this. All the literature
data consistently deviate from ours in one direction, albeit by
different amount, hinting that the variation may be real. Next, our
own data show that during the recent years the quasar is in relatively
stable state.

The $R_C$ band in the rest-frame of the quasar correspond to the UV
flux between 970 and 1420 \AA, including the Ly$\alpha$ emission line.
To compare the variability properties of Q\,2203+292 with those of
lower redshift quasars, we calculated the structure function S($\tau$),
which is commonly used to characterize the variability of large quasar
samples (i.e. Hughes, Aller \& Aller 1992):
\begin{equation}
S(\tau) =\langle [m(t)-m(t+\tau)]^2 \rangle.
\end{equation}
Here, $m(t)$ is the magnitude at time $t$ and $\tau$ is the time interval
between the two measurements in the QSO rest frame. The broken brackets
express ensemble average over measurements with the same time interval.
The structure function is less sensitive to the inhomogeneity of the
observational coverage, and it can be applied to both individual objects
and to samples of objects. Note that sometimes in the literature the
structure function is defined as a square root of the ensemble average.

\begin{figure}
\hspace*{0.0cm}\psfig{file=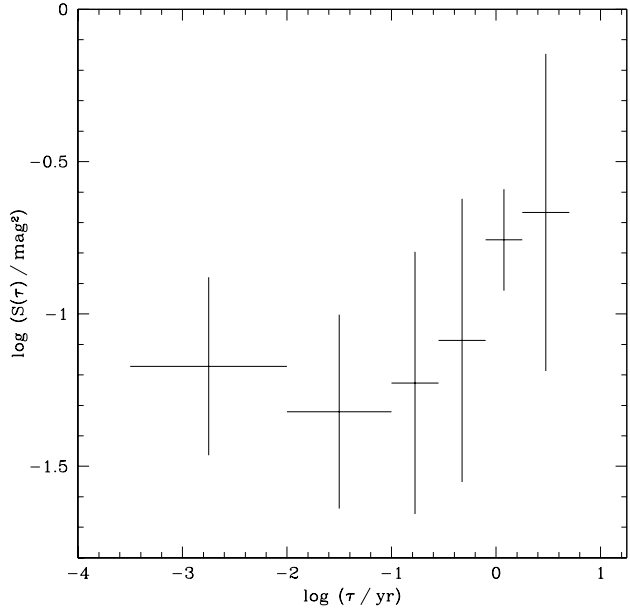,width=8.7cm,height=8.7cm}
\caption{$R$-band structure function for the quasar Q\,2203+292 in the rest
frame time. The horizontal bars mark the width of the bins used in the
ensemble averaging and the vertical ones are the corresponding rms}
\label{fig6}
\end{figure}

The structure function for Q\,2203+292 is shown in Fig.~\ref{fig6}.
The small number of measurements that form each bin lead to large
uncertainties making it difficult to draw definite conclusions. The overall
shape of S($\tau$) for Q\,2203+292 is similar to that of other QSOs
studied in the literature (i.e. Vanden Berk et al. 2004). It is
dominated by observational errors for short time intervals and by the
intrinsic QSO variability for the longer ones. There is indication
that S($\tau$) may reach a plateau at time scale just above 1 yr.
However, quasars are known to vary on a much longer time-scale, so we
interpret this as poor sampling. Note that the structure functions of
some QSOs may show intrinsic structure in S($\tau$) that is often
interpreted as variability driven by more than one physical mechanism (see
Hughes et al. 1992 for examples). We can not exclude that this can be
the case with Q\,2203+292. Further observations over longer time span
are necessary to address this question.

A quick comparisons with the literature, typically for $\tau$$\sim$0.5-2 yr,
shows that Q\,2203+292 is relatively more variable than the average
QSO in the samples of Cristiani et al. (1996) and Wold et al. (2007) and
comparable with those of di Clemente et al. (1996) and de Vries et al.
(2003).

Recently, Wold et al. (2007) explored the dependence of the quasar
variability from the mass of the central black hole. For
$M_{BH}$=10$^8$--10$^9$ M$_\odot$ -- as estimated for Q\,2203+292 by
Turner (1991) -- they obtain $R$-band amplitude of 0.3-0.4 mag, which
is similar to the value reported here.

\subsection[]{Physical properties of Q\,2203+292\label{SecPhysProp}}

To calculate the absolute luminosity of Q\,2203+292, we adopted the
following cosmological parameters: $\Omega_{\Lambda}$=0.7,
$\Omega_{M}$=0.3, and $H_0$=70 km s$^{-1}$ Mpc$^{-1}$.

The absolute $R$-band magnitude, $M_R$, is related to the apparent
magnitude, $R$, by
\begin{equation}
M_R=R-A_R-5~log~d_L-2.5~log\,(1+z)-25+\Delta\,R_{k corr}(z)
\end{equation}
\noindent
where $A_R$ is Galactic absorption, and $d_L$ is the luminosity
distance for a flat Universe.

\begin{figure}
\vspace{0.0cm}
\hspace*{0.2cm}\psfig{file=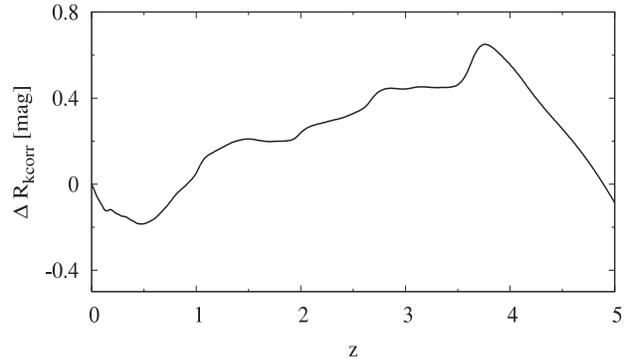,width=8.1cm,height=4.7cm}
\caption{$R_C$-band K-correction as a function of the redshift.}
\label{fig7}
\end{figure}

\begin{figure}
\hspace*{0.1cm}\psfig{file=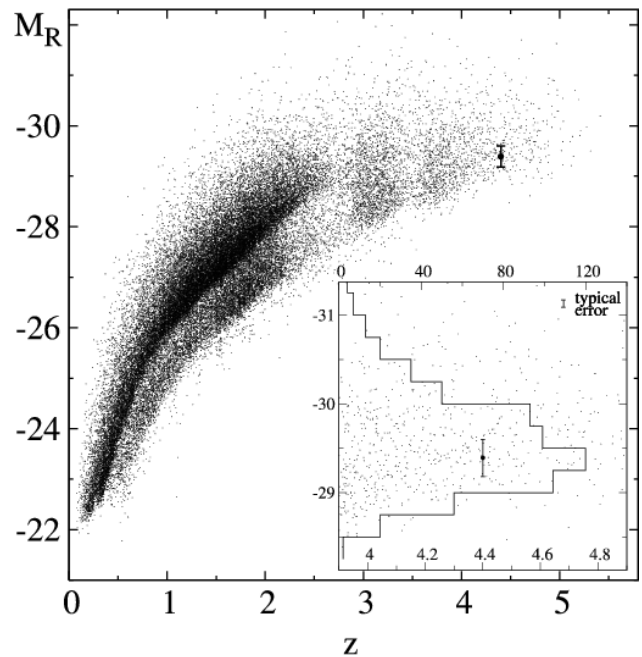,width=8.3cm,height=8.6cm}
\caption{Comparison of the absolute magnitude $M_R$ of Q\,2203+292 and
46420 SDSS quasars from Schneider et al. (2005). The apparent $R$ magnitudes
of the SDSS quasars were calculated from the SDSS $r^\prime$ magnitudes according
to the colour transformations described in Sec.~\ref{Sec_MagTransf} (see
also Fig.~\ref{fig4}, bottom panel). The bigger dot marks the average
absolute luminosity of Q\,2203+292 and the errorbars are the rms The
inset shows a histogram of $M_R$ for 625 quasars at $z$ between 3.9 and
4.9. Again, the location of Q\,2203+292 measurement is shown with a
bigger dot.}
\label{fig8}
\end{figure}

The K-correction $\Delta\,R_{k corr}(z)$ is calculated by
convolving a quasar spectrum with a sensitivity curve for a
standard $R_{C}$ filter (Fig.~\ref{fig7}). We used our
composite spectrum (see Sec. 2.4) for the SDSS quasars observed
in $r^\prime$ (Schneider et al. 2005) and the Q\,2203+292 spectrum
from Constantin et al. (2002) for our target. In the latter case
$\Delta\,R_{k corr}$=0.25 mag, 0.06 mag smaller than the value
derived for $z$=4.40 from the composite spectrum. Although this
is a systematic rather than random error, we added this difference
in quadrature to the $M_R$ uncertainty (equal to the rms of all
Q\,2203+292 measurements) to obtain a conservative
error estimate. Assuming $A_R$=0.25 mag (Schlegel, Finkbeiner \&
Davis 1998) and using the average $R_C$=20.46 mag we obtained
$M_R$=$-$29.39 mag, typical for the SDSS quasar
distribution at that redshift (Fig.~\ref{fig8}).

The minimum mass of the emitting gas in the quasar broad line region
(BLR) can be estimated following Baldwin et al. (2003), assuming Case B
recombination, when all Ly$\alpha$ photons escape and the electron
temperature is $T_e$=20,000 K:
\begin{equation}
M_{BLR} = 5.1 (10^{11}/n_e) (L_{{\rm Ly}\alpha}/10^{45})~M_{\odot},
\end{equation}
where $L_{{\rm Ly}\alpha}$ is the Ly$\alpha$ luminosity and $n_e$ is the
electron density.

We measured $L_{{\rm Ly}\alpha}$ from the spectra of
Schneider et al. (1989) and Constantin et al. (2002): 8.6$\times$10$^{43}$
and 5.6$\times$10$^{43}$ erg s$^{-1}$, respectively. Naturally, these
are only lower limits because of the Ly$\alpha$ self-absorption. In both
cases, we fitted the quasar continuum with a power law
$F_{\nu} \propto \nu^\alpha$, and fixing $\alpha$=$-$1.0. Assuming
$n_e$=10$^{11}$ cm$^{-3}$, we obtain $M_{BLR}$=0.44 and 0.29 M$_\odot$
for the measurements from the two spectra.

\subsection{Miscellaneous: Search for associated emission line objects}

Narrow band Ly$\alpha$ imaging down to 25.5 mag per sq. arcsec (McCarthy
et al. 1988) yielded no other emission line sources at the same redshift
within 2$\times$2 arcmin$^2$ from Q\,2203+292. Thompson, Djorgovski \&
Beckwith (1994) failed to find associated [O{\sc iii}] emitters in
18.2$\times$19.4 arcsec$^2$ field centred at the quasar.

We used the Photometrics\,AT200A camera at the 2m telescope at the
Rozhen Observatory to carry out a search for associated emission line
objects on 2006 August 19. The observations were obtained through a narrow
band (FWHM=32 \AA) interference filter $IF658$ centred at 6572 \AA,
corresponding to Ly$\alpha$ at $z$$\sim$4.4. Our field of view was 5
arcmin$^2$, which is much bigger than that in the earlier studies. The
exposure time was 2 h, which was split into six 1200 s exposures.
We found no evidence for sources with emission lines falling into the
bandpass of our narrow band filter down to a surface brightness level of
$\sim$24.5 mag per sq. arcsec, in agreement with the previous
attempts.

\section{Summary}

We carried out multi-year photometric $R_C$-band monitoring of the
$z$=4.40 radio quiet quasar Q\,2203+292, and we found that it exhibits
maximum peak-to-peak difference between two points on the light
curve of $\sim$0.3 mag for our data and $\sim$0.9 mag when combined
with older literature data. The rms amplitude of the lightcurve is
0.08 mag and 0.20 mag, respectively. The detected variability is at
$\sim$3$\sigma$ level when the photometric accuracy of the both data
sets are taken into account. The Monte Carlo simulation can not
reproduce the observed variation with a constant source in 10$^6$
simulations, if we consider all the data but it does in 2.5 per cent of the
simulations if we exclude the literature data and in 13.6 per cent is we
consider only the Rozhen and SAO observations. These results lead us to the
conclusion that during the recent years the quasar is in a stable state
but we refrain from making a strong statement about the earlier
variability because of possible unaccounted systematic effects in the
transformation between the different photometric systems.

Unlike previous works, which used large samples of quasars to determine
their variability properties, our goal was to assemble a well sampled
light curve of individual quasars. The structure function analysis
concluded that the object shows variability properties similar to those
of the lower redshift quasars. We also found that narrow-band imaging
at the redshifted Ly$\alpha$ shows no other emission line objects
within 5$\times$5 arcmin$^2$ field.

\section{Acknowledgments}
This work was partially supported by the following grants: $VUF201/06$,
$MUF04/05$, $F1302/03$ of the Bulgarian Science Foundation and $008/07$
with the Sofia University.
The authors thank Dr. David Crampton for giving us access to unpublished
data and to Dr. Anca Constantin for providing the Q\,2203+292 spectrum
and the median composite spectrum of high redshift quasars.
We thank Dr. Haralambi Markov for his help with the spatial correction
of the instrumental magnitudes, Dr. Tanyu Bonev for his assistance with
the FoReRo2 observations and Dr. Ilia Roussev for editing the manuscript.
We also wish to thank the anonymous referee for the useful comments that
helped to improve the paper greatly.

\label{lastpage}

\end{document}